\documentclass[conference,compsoc]{IEEEtran}
\ifCLASSOPTIONcompsoc
  \usepackage[nocompress]{cite}
\else
  \usepackage{cite}
\fi
\ifCLASSINFOpdf
\else
\fi
\usepackage{nicefrac}
\usepackage{siunitx}
\usepackage{fontawesome}
\usepackage{array,framed}
\usepackage{booktabs}
\usepackage{
  color,
  float,
  epsfig,
  wrapfig,
  graphics,
  graphicx,
  subcaption
}
\usepackage{CJKutf8}
\usepackage{fontawesome}
\usepackage{makecell}
\usepackage{textcomp}
\usepackage{setspace}
\usepackage{latexsym,fancyhdr,url}
\usepackage{enumerate}
\usepackage{algorithm2e}
\usepackage{algpseudocode}
\usepackage{graphics}
\usepackage{enumitem}
\usepackage{xparse} 
\usepackage{xspace}
\usepackage{multirow}
\usepackage{threeparttable}
\usepackage{csvsimple}
\usepackage{balance}
\usepackage{xcolor}
\usepackage{url}
\usepackage{tabularx}
\usepackage[most]{tcolorbox}
\usepackage{listings}
\usepackage{colortbl}
\usepackage{pifont}
\usepackage{titlesec}
\usepackage{tikz}
\usetikzlibrary{positioning}
\usepackage{microtype}
\usepackage[normalem]{ulem}

\titlespacing\section{0pt}{5pt plus 2pt minus 2pt}{2pt plus 2pt minus 2pt}
\titlespacing\subsection{0pt}{3pt plus 2pt minus 2pt}{2pt plus 2pt minus 2pt}
\titlespacing\subsubsection{0pt}{1pt plus 2pt minus 2pt}{2pt plus 2pt minus 2pt}

\definecolor{codegreen}{rgb}{0,0.6,0}
\definecolor{codegray}{rgb}{0.5,0.5,0.5}
\definecolor{codepurple}{rgb}{0.58,0,0.82}
\definecolor{backcolour}{rgb}{0.95,0.95,0.92}
\definecolor{qinglv}{rgb}{0,0.64,0.59}

\lstdefinestyle{mystyle}{
    backgroundcolor=\color{backcolour},   
    commentstyle=\color{codegreen},
    keywordstyle=\color{magenta},
    numberstyle=\tiny\color{codegray},
    stringstyle=\color{codepurple},
    basicstyle=\ttfamily\footnotesize,
    breakatwhitespace=false,         
    breaklines=true,                 
    captionpos=b,                    
    keepspaces=true,                 
    numbers=left,                    
    numbersep=5pt,                  
    showspaces=false,                
    showstringspaces=false,
    showtabs=false,                  
    tabsize=2
}

\lstdefinelanguage{yaml}{
    keywords={true,false,null,y,n},
    keywordstyle=\color{blue}\bfseries,
    ndkeywords={},
    ndkeywordstyle=\color{darkgray}\ttfamily,
    identifierstyle=\color{black},
    sensitive=false,
    comment=[l]{\#},
    morecomment=[s]{/*}{*/},
    commentstyle=\color{magenta}\ttfamily,
    moredelim=[is][\textcolor{qinglv}]{\%\%}{\%\%},
    moredelim=**[is][\color{blue}]{\$\$}{\$\$},
    escapeinside={(*@}{@*)}, 
}

\usepackage{
  tikz,
  pgfplots,
  pgfplotstable
}
\usepackage{hyperref}

\usetikzlibrary{
  shapes.geometric,
  arrows,
  external,
  pgfplots.groupplots,
  matrix
}

\pgfplotsset{compat=1.9}

\usepackage{mathtools,}

\DeclareMathAlphabet{\mathcal}{OMS}{cmsy}{m}{n}

\DeclareGraphicsExtensions{%
    .png,.PNG,%
    .pdf,.PDF,%
    .jpg,.mps,.jpeg,.jbig2,.jb2,.JPG,.JPEG,.JBIG2,.JB2}

\begin{document}
%

\title{Investigating Security Implications of Automatically Generated Code on the Software Supply Chain}

\author{
\IEEEauthorblockN{Xiaofan Li}
\IEEEauthorblockA{University of Delaware\\
Newark, Delaware\\
xiaofan@udel.edu}
\and
\IEEEauthorblockN{Xing Gao}
\IEEEauthorblockA{University of Delaware\\
Newark, Delaware\\
xgao@udel.edu}}



%


\maketitle

\begin{abstract}
In recent years, various software supply chain (SSC) attacks have posed significant risks to the global community. 
Severe consequences may arise if developers integrate insecure code snippets that are vulnerable to SSC attacks into their products. 
Particularly, code generation techniques, such as large language models (LLMs), have been widely utilized in the developer community.
However, LLMs are known to suffer from inherent issues when generating code, including fabrication, misinformation, and reliance on outdated training data, all of which can result in serious software supply chain threats. 
In this paper, we investigate the security threats to the SSC that arise from these inherent issues. 
We examine three categories of threats, including eleven potential SSC-related threats, related to external components in source code, and continuous integration configuration files. 
We find some threats in LLM-generated code could enable attackers to hijack software and workflows, while some others might cause potential hidden threats that compromise the security of the software over time. 
To understand these security impacts and severity, we design a tool, SSCGuard, to generate 439,138 prompts based on SSC-related questions collected online, and analyze the responses of four popular LLMs from GPT and Llama. 
Our results show that all identified SSC-related threats persistently exist. 
To mitigate these risks, we propose a novel prompt-based defense mechanism, namely Chain-of-Confirmation, to reduce fabrication, and a middleware-based defense that informs users of various SSC threats. 
\end{abstract}


%
\IEEEpeerreviewmaketitle

\label{sec:introduction}

\section{Introduction}

The open-source software supply chain (SSC) has been an integral part of today’s software development.
Software companies largely rely on open-source components for software development.
Millions of packages are actively maintained in various software registries, attracting trillions of downloading requests per year~\cite{SonatypeReport}.
Additionally, many organizations have adopted continuous integration (CI) services to automate the building and testing of their code changes.
Thus, vulnerable or insecure elements
in SSC can cause serious security threats with potentially massive financial loss.
Unfortunately, in recent years, various SSC attacks have posed significant risks to the global community. 
One example is \textit{ua-parser-js}, a popular npm package with 7 million weekly downloads, which was hijacked to deliver malware in 2021, affecting numerous users~\cite{ua-parser-js}.

One reason that vulnerabilities are introduced during software development is the availability of a large amount of ready-to-use, yet insecure, code snippets on the Internet.
Not only inexperienced programmers,  but a large part of the developer community tend to simply copy\&paste ready-to-use code snippets into production software~\cite{fischer2017stack}. 
While the rich programming source provides quick solutions and enables fast prototyping, such code may not always be secure and could contain vulnerabilities. 
For example, previous research has observed millions of Android applications contain insecure code snippets directly copied from Stack Overflow~\cite{fischer2017stack}.

The recent advent of large language models (LLMs), such as ChatGPT, further exacerbates the situation. 
By providing an input prompt describing a code-related problem, LLMs can immediately generate corresponding code. 
Owing to their efficiency, LLMs have been widely utilized to generate code by both novice and expert programmers~\cite{cui2024effects}. 
However, research has demonstrated that LLM-generated code is prone to have security vulnerabilities~\cite{pearce2022asleep,sandoval2023lost}. 

Particularly, LLMs are known to suffer from several inherent issues. 
One such issue is fabrication, where the generated content appears plausible but is in fact non-existent.
Another is misleading, where the output seems accurate and trustworthy but contains factual errors. 
In addition, because LLMs are trained on large static data, they are inherently prone to producing outdated information. 
Unfortunately, such inherent issues could cause serious SSC-related security threats, potentially enabling attackers to hijack software/system relying on the generated code. 
One example is the \textit{package hallucination}~\cite{spracklen2024we} threat: the generated code might reference an external but non-existent package that adversaries can preemptively create and exploit. 
\textit{In parallel with our work, Spracklen et al.~\cite{spracklen2024we} independently explored the package hallucination and appeared online earlier. }
However, \textit{package hallucination} is only one of the many threats within the broader scope of SSC threats that can be introduced by the generated code. 
Despite the extensive recent works have studied both the security issues in generated code~\cite{liu2024exploring,liu2024no,zhang2024effectiveness}, there is no systematic study toward investigating the threats to the SSC posed by automatically generated code.

In this paper, we present a systematic study to understand the potential threats of generated code across different phases of the software supply chain. 
We examine three inherent issues of LLMs, including fabrication, misleading, and reliance on outdated training data. 
Then, we investigate the SSC threats that these issues might introduce into source code and CI configuration files. 
Specifically, we find that generated code contains not only hallucinated packages, but also hallucinated domains, libraries, GitHub accounts, CI plugins, and other third-party resources. Many of these non-existent components could be exploited by adversaries to preemptively take ownership and hijack following users. 
In addition, we find that the generated code could reference outdated, deprecated, or even vulnerable external components. For example, LLMs might reference CI plugins with versions that are vulnerable to version reuse attacks~\cite{li2024toward}, or generate CI configuration files containing code injection vulnerabilities~\cite{muralee2023argus}.

To investigate the existence of the aforementioned threats and assess their severity, we design a tool, SSCGuard, and conduct a large-scale analysis on four widely-used LLM models selected from the GPT and Llama series.
The idea is to simulate how normal developers interact with LLMs when seeking code-related assistance. 
Basically, we collect SSC-related questions from Stack Overflow and transform them into different prompts, asking LLMs to provide source code and CI configuration files. 
Our analysis covers packages of five programming languages (i.e., Nodejs, Python, Ruby, PHP, and Perl), third-party resources including JavaScript libraries and CSS files, and GitHub Actions (CI service).
In total, we have constructed over 400k prompts. 

The experimental results show that all threats widely exist in the generated code for all targeted LLMs.
For example, we find that hallucinated packages, domains, and GitHub accounts consistently occur for all programming languages.
Also, the performance drops significantly (i.e., more hallucinated contents) if we explicitly ask LLMs to recommend external components. 
Most concerningly, LLMs perform poorly in recommending CI plugins, with the majority being hallucinated.
For the rest of them, LLMs also have problems with their versions, such as recommending hallucinated versions or versions that are vulnerable to version reuse attacks.
In addition, LLMs frequently recommend vulnerable external components in the generate code, some of which contain high-severity security vulnerabilities.
Finally, LLMs often provide outdated external components, some of which have been removed or deprecated. 
Others are redirected to new locations, enabling attackers to hijack software that depends on the original references.

As mitigation, our proposed tool can be easily deployed as an intermediary layer between the LLM and developers, to identify problematic external components in LLM-generated code and inform developers.
We further propose a novel 
prompt-based defense, where LLMs are instructed to verify the existence of the packages referenced in their generated code, and then re-generate the code.
Results show that our method can reduce the package hallucination rate to half, while providing a similar number of packages.

\label{sec:background}

\section{Background}

\subsection{Software Supply Chain}

The software supply chain (SSC) involves a series of third-party components and processes that collectively contribute to software development. Below, we briefly introduce several key phases.

\noindent{\textbf{Development.}}  
Software source code is typically integrated with various external components. 
There are several commonly utilized external components: 
(1) \textit{Packages} are distributable units of code that are widely utilized by developers to accelerate software development. 
Packages are usually hosted on registries, which serve as platforms for developers to publish and share their packages.
In this work, we focus on five package registries corresponding to different programming languages: Nodejs (npm~\cite{npm_registry}), Python (PyPI~\cite{pipy_registry}), PHP (Packagist~\cite{packagist_registry}), Ruby (RubyGems~\cite{rubygems_registry}), and Perl (CPAN~\cite{cpan_registry}). 
(2) \textit{Third-party resources/services}. 
A variety of other third-party resources/services are also widely utilized at this phase, including JavaScript libraries, CSS files, and various APIs provided by different service providers.  
These resources can be accessed directly via URLs. 
For example, a JavaScript library can be included by referencing its URL in the \texttt{src} attribute of a \texttt{<script>} tag without the need to install it through a package registry (i.e., npm).

\noindent{\textbf{Build.}}
This phase focuses on compiling and transforming the source code 
into the final software product.
To date, many organizations rely on Continuous Integration (CI) to integrate and build code efficiently. 
Typically, CI platforms (e.g., GitHub Actions) execute the repository's CI workflows, which define a series of tasks, within a configuration file (i.e., \texttt{YAML} files). 
Since many tasks are repetitive, sharing them across CI communities promotes reuse and efficiency. 
To support this, CI platforms introduce plugins to accelerate workflow development. 
Developers can reference a plugin by specifying its location (e.g., the repository hosting the plugin) and the desired version. 
Once the workflow is triggered, the specified plugins will be retrieved and executed as part of the CI procedure.

\noindent{\textbf{Distribution.}} This phase involves distributing the software to downstream users, such as through marketplaces (e.g., package registries). 
Notably, a package is also a form of software. 
Other developers can then search for, install, and utilize it in their software.

\subsection{External Components Integration}

Since external components vary in type (e.g., packages and plugins), their integration comes with different requirements.

\noindent{\textbf{Packages.}} 
To integrate a package into the source code, developers typically begin by installing it through a package command-line tool (e.g., \texttt{pip}). 
For example, a Python package can be installed using the command \texttt{pip install "package-name"}, where \texttt{pip} is a command-line tool for installing and managing Python packages. 
By default, this command installs the latest available version. 
Developers can also specify package versions during installation, as in \texttt{pip install "package-name==1.4"}, where \textit{1.4} indicates the desired version. 
Once installed, developers can utilize the installed packages to support further development within the source code.

\noindent{\textbf{Third-party resources/services.}}
These are often integrated into source code by referencing their URLs. 
Unlike packages, they do not require installation via command-line tools and can be accessed directly through their URLs. 
Such resources/services are hosted on servers identified by the domain names in the URLs. 
Versioning is also commonly supported and specified as part of the URL. 
For example, \texttt{https://example.com/libs?v=1.2.3}, where \texttt{example.com} is the domain, \texttt{1.2.3} indicates the version. 
Additionally, these resources/services can also be hosted on third-party platforms, such as Content Delivery Networks (CDNs) links and GitHub repositories. 
GitHub repositories can be utilized to provide "code as a service". 
For example, developers can host APIs or libraries within a GitHub repository, allowing them to be directly referenced and utilized by other developers. 
A remote GitHub repository can be accessed through the URL \texttt{https://github.com/\{user\}/\{repo\}}~\cite{remote_repository}, where \texttt{user} represents the GitHub account.

\noindent{\textbf{CI configuration.}}
The tasks of a CI workflow are defined within a CI configuration file, which is typically written in \texttt{YAML} format. 
These configuration files can be integrated into one another, allowing a workflow to be reused across multiple CI configurations~\cite{li2024toward}.

\noindent{\textbf{CI Plugins}}
Plugins integrated within the CI configuration files are typically hosted on GitHub repositories. 
To reference a plugin, developers use the format \texttt{user/repo@ref}, where \texttt{user} is the GitHub account, \texttt{repo} is the repository where the plugin rests, and \texttt{ref} specifies the version of the plugin, which can be a tag, branch, or commit hash.

\subsection{SSC Threats}
Security threats targeting the SSC have become prevalent in recent years.
One popular attack vector is to exploit dangling references, including dangling domains and dangling packages. 
A dangling reference, similar to a dangling pointer in the use-after-free vulnerability, refers to an expired or non-existent package/domain that is still referenced by others (e.g., users or software). 
Attackers can hijack the dangling reference, such as by publishing a new package under the same name, or by registering the expired domain. 
After that, attackers can further inject malicious code into the package or manipulate resources associated with the domain. 

Meanwhile, the widespread adoption of various CI tools also exposes new attack surfaces, potentially enabling attackers to inject malicious code during the \textit{build} stage.
As one of the top five challenges in SSC security~\cite{enckTopFiveChallenges2022}, many security incidents~\cite{jarlobGHSL2020235ArbitraryCommand2021,goodinHackersBackdoorPHP2021,hanleySecurityAlertAttack2022} have happened in recent years.
For example, previous research~\cite{gu2023continuous,gu2024more} has demonstrated that the weak isolation among CI components may allow users to access unauthorized resources.

Finally, version reuse is a known threat in package management~\cite{gu2023investigating}.
A reused version (i.e., different packages with the same version number) enables attackers to inject malicious code while keeping the version number unchanged, and thus users might not notice such a modification. 
While most software registries have disabled version reuse~\cite{NpmunpublishNpmDocs,stufftDistutilsClosingDelete2015}, researchers have demonstrated that version reuse still occasionally happens in some registries~\cite{gu2023investigating}.
Even worse, third-party CI plugins are also vulnerable to version reuse~\cite{li2024toward}, as their version control relies on git tags or git branch names, which could be easily deleted and then reused.

\subsection{LLMs for Code Generation and Hallucinations}

Large Language Models (LLMs) have achieved significant advancements in a wide range of tasks, including code generation.  
Instead of writing code line by line, developers can now describe their problems or requirements in natural language, allowing LLMs to generate working code based on these descriptions. 
LLMs capable of generating code are trained on extensive datasets comprising code repositories (e.g., GitHub), programming forums (e.g., Stack Overflow), and other web-based resources related to coding.

\noindent\textbf{LLM Hallucinations}  refer to the phenomenon that generated text is either factually incorrect, nonsensical, or diverges from the inputs. 
Hallucinations in code generation can cause severe consequences, such as the spread of misinformation and potential legal liability~\cite{hao2024quantifying}.
However, most of the research on hallucinations remains within the context of the generated code itself~\cite{liu2024exploring,liu2024no,zhang2024effectiveness,jesse2023large, li2024enhancing, tambon2024bugs}. 
In the broader scope of the SSC, there exists a range of third-party components that extend beyond the code alone. 
LLMs could generate hallucinated content that appears normal and secure from the standard code practices, but is actually vulnerable to SSC threats.  
Such hallucinations in LLM-generated code are more covert, potentially leading to unexpected security vulnerabilities.

\begin{figure}[t]
\includegraphics[scale=0.6]{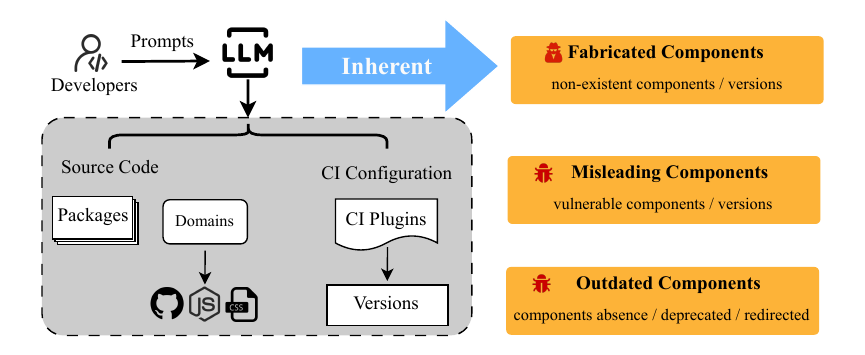}
\caption{Overview of Threats in SSC.}
\vspace{-2mm}
\label{fig:threat_model}
\end{figure}

\section{Security Model and Principles}
\label{sec:security_model}

This paper investigates potential security vulnerabilities in the code generated by LLMs, focusing on how such vulnerabilities could be exploited to carry out various SSC attacks (e.g., component redirection hijacking, and distributing malicious versions of legitimate components). 
Our goal is to systematically investigate the spectrum of potential SSC threats embedded in LLM-generated code. 
In the following, we first introduce the threat model and then discuss the security principles inherent in LLMs that might inadvertently introduce or amplify different types of SSC threats.

\subsection{Threat Model}

We generally consider a typical software development scenario in which developers utilize LLM to assist with code writing. 
We focus on the code related to SSC, such as code generated during the \textit{development} and \textit{build} phases. 
We consider victims to be typical software developers, especially inexperienced developers with less security awareness. 
As LLM users, they provide plain natural language descriptions (i.e., prompts) of the desired functionality or problem that needs to be solved, and the LLMs automatically generate corresponding code. 
With limited security knowledge in SSC, these developers might simply utilize and integrate LLM-generated code into their software without verifying its safety.  

We assume the LLMs are benign and not compromised. 
Developers either use well-established commercial LLMs (e.g., ChatGPT) or open-source LLMs (e.g., Llama~\cite{llama}).
Except for their inherent properties (e.g., hallucination), these LLMs do not contain vulnerabilities such as backdoor or poisoned training data~\cite{aghakhani2024trojanpuzzle, yan2024llm, schuster2021you}. 
We also assume the usage of LLMs is standard and secure.

In particular, we consider two scenarios of using LLMs that are related to SSC. 
Figure~\ref{fig:threat_model} illustrates the threat overview. 
For the first scenario, developers request a source code related problem with the LLMs, which then generates a response containing functional code aimed at solving the issue. 
The generated code may include/import external components, such as packages, JavaScript libraries, CSS files, and other third-party resources/services (e.g., APIs). 
These components are often suggested to enhance functionality or streamline development. 
In addition to the generated code, the response might also include related references, such as GitHub repositories, that provide additional context or information for the solution. 
In the second scenario, developers turn to LLMs for help in building their projects by providing a description of the desired CI functionality.
In response, the LLM generates a CI workflow configuration file (typically in YAML format) based on the given specifications.
This generated configuration may also include references to specific CI plugins, along with their corresponding versions, to support the defined tasks.

For adversaries, we assume that they are just normal LLM users without the capabilities to modify these well-established LLMs. For example, they cannot inject malicious code payloads into the LLM's training data or modify internal (hyper)-parameters of LLMs.
As a result, the adversaries cannot provide designated triggers that would prompt LLMs to suggest vulnerable code. 
They also cannot inject prompts into LLM-integrated applications or manipulate the returned response to victims.
Instead of breaking the target LLMs, the adversaries are also legitimate users of LLMs. They provide various prompts and record the corresponding responses. 
To further exploit the vulnerabilities described in this paper, adversaries can continuously monitor various key events (e.g., package existence) in third-party platforms and other external resources (e.g., domains). 
Such information can be easily collected from public sources, such as crawling or querying corresponding APIs. 
They can also legitimately take control of some resources, such as registering an account, publishing a package, or registering a domain.

\subsection{Fabricated External Component Information}
LLMs can generate content that appears plausible but is factually incorrect in natural language processing tasks. 
Similarly, such fabrication can also arise in external components when LLMs are used to generate code. 
In such cases, the models might generate code that references components that appear to be valid but do not actually exist. 
In addition, the generated code often includes detailed usage examples, such as specifying which functions to call and what types of parameters to provide. 
These seemingly coherent examples can easily lead developers to believe that the component is legitimate, without realizing it might have been fabricated.

In some cases, developers include specific constraints in the prompts they provide to LLMs that reflect the requirements of their code. 
For example, the code might need to be compatible with an older version of a system or platform. 
To satisfy these constraints, the LLM might generate code that references a particular version of a component, and likewise, the LLMs provide code examples demonstrating how to use the specified version of the component. 
These convincing details can lead developers to trust that the component and the specified version exist, even when they do not. 
If LLMs suggest a valid component along with a version that does not actually exist, it is possible for the fabricated version to be later published and made available, giving the appearance of legitimacy despite not originating from the original development process.

LLMs might fabricate external components in different ways depending on the type of component. 
In the case of packages, the fabrication is often direct and simple. 
The models might generate package names that do not actually exist. 
Once these names appear in the generated code, they become available, and any party (including adversaries) can claim and publish packages under those names.

For third-party resources/services, these resources and services are typically hosted on servers linked to the domains specified in the provided URLs. 
If LLMs generate a third-party resource/service URL whose domain does not exist, then any party might register or purchase the domain in advance and host the resource/service under that domain. 
For third-party resources/services hosted on third-party platforms (i.e., CDNs and GitHub repositories), LLMs may fabricate specific components of the URLs. 
This includes non-existent package names and versions in CDN URLs, or fabricated GitHub account names in repository URLs. 
These fabricated components can be taken in advance by registering the corresponding package names or accounts.

For CI plugins, the situation mirrors that of GitHub repositories. 
If LLMs suggest a valid plugin (i.e., GitHub repository) with a fabricated version (e.g., a fabricated tag), that version can be easily created to appear legitimate. 
For example, a pull request can be submitted to add the suggested tag.
CI plugins support both branches and tags that share the same name (known as mixed reuse~\cite{li2024toward}). 
A branch with the reused name can also be created through a pull request, resulting in the branch version being used by default.

\subsection{Misleading External Components Information}

Another well-documented behavior of LLMs in natural language is the generation of misleading content. 
This occurs when the output appears accurate and trustworthy but contains subtle ambiguities or factual errors that might lead to misunderstanding. 
A similar pattern can emerge in code generation, where LLMs may reference external components that are valid but known to contain vulnerabilities (i.e., vulnerable components). 
This includes vulnerabilities in both the external components themselves and their versions specified by the models.

Each type of external component is associated with a corresponding version control mechanism. 
For example, npm packages typically follow semantic versioning~\cite{npm_versioning}, while CI plugins rely on git-based versioning~\cite{li2024toward}, which includes branches, tags, and commit hashes.
One of the most important reasons for using version control is to keep components secure. 
Since components are essentially code, developers regularly find vulnerabilities in components and release new versions to fix them. 
Versions that contain vulnerabilities are referred to as vulnerable versions. 
When LLMs generate code, they might reference valid components, but the specified versions are vulnerable. 
Since the component is valid and accompanied by seemingly correct code examples, developers might trust it without further verifying the specified version.

GitHub Actions supports reusable workflows, allowing one CI configuration to reference another. 
As a result, CI configurations can themselves be treated as external components within the \textit{build} process. 
Vulnerable CI configurations can introduce unintended behavior during the \textit{build} process, increasing the risk of misuse or exploitation with the SSC.

\subsection{Outdated External Components Information}

Since LLMs are trained on large static data that captures a snapshot of the web at a given point in time, they are inherently prone to generating outdated information. 
This limitation can manifest in code generation, where LLMs might suggest outdated external components. 
Several common scenarios illustrate how outdated components can be referenced in the generated code.

The most common scenario is that external components are removed, making them inaccessible. 
LLMs might still generate code that references these components, which no longer exist due to various reasons. 
This typically occurs with third-party resources/services, and differs from fabrication in that the domain hosting the resources/services remains valid.

Another common scenario involves external components that are still accessible but no longer maintained, indicating they have been deprecated. 
While these components remain valid, their use can lead to hidden risks. 
For example, as software evolves, these components might no longer be compatible with other components within the software.

The last scenario occurs when the owners of external components change their organization names, resulting in redirection from the original addresses to new ones. 
For example, in the case of third-party resources/services hosted on GitHub repositories and CI plugins, accounts might be changed, and GitHub automatically redirects requests from original accounts to the new ones. 
However, this redirection leaves the original accounts available for re-registration. 
If the original accounts are re-registered by new parties, the redirection will be broken, potentially causing any software that references components under the original accounts to resolve to unintended sources.

\section{Identified Threats}
\label{sec:identified_threats}

Based on the threat model and security principles, we identify several potential SSC security threats that might be introduced by LLM-generated code. 
These threats might arise when the generated code references external components that are fabricated, misleading, or outdated.
Such references can lead to scenarios where these components are unintentionally exposed to hijacking or manipulation, potentially compromising the security and reliability of the software that integrates them.

\subsection{External Component Hallucination}

External component hallucination refers to the phenomenon where LLMs fabricate components that do not actually exist. 
These fabricated components can be registered and controlled by adversaries, potentially leading to the compromise of any software that relies on them. 

\subsubsection{Package Hallucination}

Non-existent package names generated by LLMs can be identified and registered in public package repositories.
Non-existent versions might be published.
For example, adversaries can contact the original package owner and request maintainer access, then publish the version as if it were legitimate. 
When developers rely on LLM-generated code that references these malicious packages, they might unknowingly install and integrate them into their software. 
Some research has focused on package hallucination~\cite{spracklen2024we, ai_hallucination_package_risk, ai_package_hallucination}. 
However, we find that it represents only one type of threat within the broader category of external component hallucination. 
Other hallucinated components, beyond packages, pose significant security risks. 

We further identify that package hallucinations exist in CDN links. CDNs, such as jsDelivr~\cite{jsDelivr} and UNPKG~\cite{UNPKG}, provide a convenient mechanism for accessing static files directly from npm packages and GitHub repositories. 
For example, a developer can request a static JavaScript file from an npm package using an UNPKG link\footnote{(e.g., \texttt{https://unpkg.com/:package@:version/:file})}. 
This package has been cached in data centers around the world. 
Therefore, CDNs (e.g., UNPKG) return the JavaScript file from the cached package. 
In such scenarios, LLMs might fabricate references to non-existent packages under valid CDN links. 
We find that such non-existent packages can also be hijacked by adversaries. 
Adversaries can first identify a non-existent Nodejs package, then publish the malicious package under the same name on npm. 
When developers reference files using the CDN link, the CDN will retrieve and cache these files from the malicious package, which are then available to developers. 

\subsubsection{Domain Hallucination}
Similar to package hallucination, LLMs can also fabricate non-existent domain names when generating code that references external JavaScript libraries, CSS files, or third-party resources/services. 
These resources are typically referenced via URLs for the hosted domains.
If adversaries identify such hallucinated, non-existent (or expired) domains, they can register and take control of them, hosting malicious content (e.g., JavaScript libraries).
Developers relying on LLM-generated code that references these domains may unknowingly introduce malicious code into their software, as adversaries can manipulate the hosted components (e.g., JavaScript libraries) to serve their malicious intent. 
For example, previous research has demonstrated that CSS files can be exploited to inject malicious code and load malware~\cite{css_injection}. 

\subsubsection{GitHub Account Hallucination}
When LLMs generate code, they might provide references to GitHub repositories that serve as hosted APIs or libraries. 
However, some of the associated GitHub accounts might not exist. 
In this case, adversaries can identify non-existent accounts and register them to gain control over the referenced repositories. 
In the end, developers relying on these references might inadvertently interact with repositories controlled by adversaries, posing security risks.

\subsubsection{CI Plugin Hallucination}

CI plugins are hosted in dedicated GitHub repositories, allowing developers to reference them directly in their CI workflows.
When a workflow is triggered, the referenced plugins are downloaded to the server executing the workflow tasks. 
However, if a plugin is unavailable because its repository does not exist, the download process will fail, leading to the failure of the entire workflow. 
In LLM-generated configuration files, LLMs might reference non-existent plugins. 
Adversaries can identify and register these fabricated plugins, taking complete control over them. 
As a result, developers, who implicitly trust the LLM-generated configuration files, might unknowingly reference such plugins, enabling attackers to introduce malicious code into their workflows.

\subsubsection{CI Plugin Version Hallucination}

When referencing a plugin in a CI workflow, it is strongly recommended to specify the plugin's version. 
The version can be specified via a Git reference (e.g., a branch or tag). 
If the version does not exist, the CI workflow fails immediately. 
For example, to reference the plugin \texttt{actions/checkout} in a workflow, developers can use the \texttt{uses} syntax, such as \texttt{uses: actions/checkout@v4.0.1}. 
If the version does not exist, GitHub Actions returns an error "\textit{unable to find version @v4.0.1}" immediately. 
LLMs might provide a non-existent version (i.e., version hallucination) when generating workflow configuration files. 
Adversaries can exploit this by creating a fabricated version through various means. 
For example, they can submit a pull request to the plugin’s repository to create the specified version (e.g., branch or tag), making it appear legitimate. 

\subsubsection{CI Plugin Version Reuse}

Previous research has demonstrated that it is problematic to reference a plugin version using its branch or tag~\cite{li2024toward}. 
In GitHub Actions, branches and tags can be deleted and recreated, allowing malicious plugin maintainers to modify the code associated with a specific version without changing the version (i.e., version reuse).
Specifically, GitHub Actions allows the creation of branches and tags with the same names. 
In such cases, GitHub Actions prioritizes the branch version by default. 
This introduces another security risk, enabling malicious plugin maintainers to create a branch with the same name as an existing tag, resulting in a branch-tag mix reuse attack~\cite{li2024toward}. 
In both cases, when LLMs generate configuration files that use Git references (such as branches or tags) for version control, they may inadvertently introduce security risks, such as referencing a version that contains malicious code.

\lstset{
  language=yaml,
  basicstyle=\ttfamily\footnotesize\selectfont,
  numbers=left,
  numberstyle=\tiny\color{codegray},
  stepnumber=1,
  numbersep=5pt,
  showspaces=false,
  showstringspaces=false,
  showtabs=false,
  tabsize=2,
  captionpos=b,
  breaklines=true,
  breakatwhitespace=true,
  breakautoindent=true,
  linewidth=0.99\linewidth,
  basewidth=0.5em,
  xleftmargin=10pt,
  lineskip=0.3pt,
  literate={'}{{\textquotesingle}}1
}
\vspace{0mm}
\begin{lstlisting}[caption={Example of Code Injection in a CI workflow.\vspace{-4mm}}, label={lst:code-injection-demo}, abovecaptionskip=5pt, float=t]
%%steps%%:                    
  - %%name%%: List Pull Request Details 
    %%run%%: echo "Title: ${{ github.event.pull_request.title }}"
# Based on a malicious input (i.e., ";id;echo "), the code that is actually executed is as follows
%%run%%: echo "Title: ";id;echo ""
\end{lstlisting}

\subsection{Vulnerable External Components}

LLM-generated code might reference external components that contain known vulnerabilities.
These might be vulnerable due to vulnerabilities in the referenced versions or inherent weaknesses in the components themselves.
Any software that integrates these components is potentially exposed to security risks.

\subsubsection{Vulnerable Third-party Resources/Services Versions}

When LLMs generate code, they might recommend third-party resources/services without guaranteeing that the suggested versions are up-to-date or free from known vulnerabilities. 
Incorporating such components might expose the software to exploitable threats. 
For example, external JavaScript libraries are heavily used in modern web applications. However, study~\cite{lauinger2018thou} shows that 37\% of websites rely on at least one JavaScript library whose version contains known vulnerabilities. 
Therefore, using a vulnerable version of a JavaScript library exposes web applications to risks, as adversaries can simply visit websites to determine whether a known-vulnerable JavaScript library is in use.

\subsubsection{Vulnerable CI Configuration}

This vulnerability arises from misconfigured workflows~\cite{muralee2023argus}, where inadequate input sanitization allows adversaries to inject malicious code into the CI workflow. 
Listing~\ref{lst:code-injection-demo} illustrates an example.
The \textit{run} step (line 4) is particularly vulnerable, as it takes input from GitHub ("\texttt{\${{github.event.pull\_request.title}}}"). 
This input could be exploited by a malicious pull request initiator. 
For example, an adversary could craft a pull request title containing the string \texttt{";id;echo~"}. 
During the workflow execution, it actually executes \texttt{echo~"Title:~";id;echo~""}, allowing the adversary to execute arbitrary commands. 
Thus, it is crucial to examine whether LLM-generated configuration files are subject to such vulnerability.

\begin{figure}[t]
   \centering
   \includegraphics[width=0.5\textwidth]{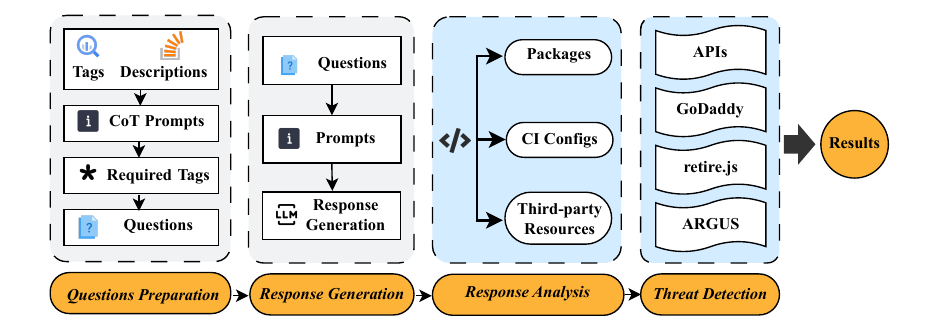}
   \caption{Overview of the SSCGuard.}
   \vspace{-2mm}
   \label{fig:overview_framework}
 \end{figure}

\subsection{Outdated External Components}

LLM-generated code may reference external components that have been removed, deprecated, or are no longer maintained. 
In some cases, a component may be redirected to a new location, leaving the original location available to be hijacked by adversaries. 
Software that integrates such components is potentially exposed to hidden security risks. 

\subsubsection{Third-party Resources/Services Absence}

In some cases of third-party resources/services, the hosted domains are valid, but the hosted resources/services do not exist. 
Adversaries cannot preemptively publish these resources, as the domains are not available for malicious takeover.
We refer to these cases as the absence of third-party resources/services, which can still lead to unexpected issues such as broken features.
Furthermore, this absence might disrupt the dependency chain, affecting other components that rely on it indirectly. 

\subsubsection{Package Deprecation}

A package registry (e.g., npm) is a central repository where packages are published and managed. 
When a package is no longer maintained by its owner, it is often marked as deprecated or abandoned, signaling to users that it should no longer be used.
Packages can be deprecated for various reasons, such as package retirement~\cite{laravelcollective_deprecated}, which leaves them without updates or bug fixes. 
Incorporating deprecated packages into software can pose significant challenges for developers, such as being unable to address arising bugs or facing compatibility issues with newer software versions. 
When LLMs generate code, they might reference deprecated packages. 
Without thorough scrutiny, developers might inadvertently integrate these packages into their software, as these packages are valid for installation. 
Since deprecated packages often remain accessible rather than being removed, their use can introduce latent security vulnerabilities, potentially compromising the software's integrity over time. 

\subsubsection{CI Plugin Redirection Hijacking}

Plugins for GitHub Actions are stored on GitHub and distributed via GitHub repositories. 
Changes to the GitHub account hosting a plugin, such as account renames or deletions, might not be immediately apparent to developers. 
As a result, developers might continue referencing the plugin's old location (i.e., the original account). 
To address this, GitHub automatically creates a redirection linking the old and new locations, enabling both references to remain functional. 
However, this redirection introduces a critical vulnerability~\cite{li2024toward}. 
If the old location becomes available for registering (e.g., account deletion), adversaries can hijack it by registering this account and publishing a new (malicious) plugin with the same name. 
The redirection link will then be broken, and developers referencing the plugin's old location may use the malicious plugin controlled by the adversaries without realizing it. 
Compared to plugin hallucination, attackers do not need to hijack the original plugins (which are stored in the new location/account). 
When LLMs are adopted to generate workflow configuration files, they might reference the old location of redirected plugins, potentially introducing risks.

\section{Methodology}
\label{sec:methodology_new}
To detect the identified SSC vulnerabilities introduced by LLM-generated code, we design SSCGuard, a tool for automated large-scale LLM-generated code vulnerability detection. 
Overall, SSCGuard first collects real-world coding questions from \textit{Stack Overflow} and transforms them into prompts. 
Then, SSCGuard feeds prompts to the target LLMs to get responses. 
Finally, SSCGuard analyzes the responses and detects the identified vulnerabilities. 
The overview is illustrated in Figure~\ref{fig:overview_framework}.

\subsection{Target LLMs}
We choose two commercial models and two open-source models as our target models. 
OpenAI's GPT series is among the most popular and high-performing commercial LLM models~\cite{lmarena}. 
The series includes a range of models, providing capable yet cost-efficient options for developers and researchers. 
We select GPT-4o-mini and GPT-3.5 Turbo, two popular choices when the experiments were conducted. 

Meanwhile, Llama is a family of open-source AI models released by Meta AI~\cite{llama}. 
When the experiments were conducted, Meta AI has released Llama 3.1. 
Thus, we choose two models from Llama 3.1.
In particular,  we select the Llama-3.1-8b-instruct model (referred to as Llama-3.1-8b for short), which is a lightweight, ultra-fast model that can be run in resource-constrained settings. 
We also use a customized model (i.e., Llama-3.1-sonar-small-128k-chat, referred to as Llama-3.1-sonar for short) created by Perplexity~\cite{perplexity}. 
Based on their claim, Llama-3.1-sonar supports up to 128K tokens, and it is ideal for chatbots and virtual assistants with fact-based answers~\cite{sonaronline}.

\subsection{Question Preparation}

\textit{Stack Overflow} is one of the most popular online platforms for programmers to ask and answer coding questions. 
The questions are categorized by tags representing different topics. 
Currently, there are over 60,000 tags and 2 million questions. 
Thus, it is a perfect place for us to collect coding questions. 
We first filter the target tags, then select the top k questions with the most answers from each tag to construct the dataset.

To filter the target tags, SSCGuard employs a Chain-of-Thought (CoT) reasoning-based prompting strategy to narrow the tags of interest. 
It begins by collecting tags from BigQuery~\cite{bigquery}, a public dataset that provides tag names and question counts from \textit{Stack Overflow}. 
Next, it obtains the tag descriptions from the \textit{Stack Overflow} website using the collected tag names. 
Then, CoT reasoning is applied to the tag names and descriptions to identify the target tags.

\begin{figure}[t]
    \centering

    \begin{tcolorbox}[boxrule=0pt, colback=blue!10]

    \footnotesize{\textbf{Q1:} 
    Give me an answer of yes or no, does this statement describe a technology used in modern application development and building: "\{tag\_name\}, \{description\}"}

    \end{tcolorbox}

    \vspace{-2mm}

    \begin{tikzpicture}
        \draw[->, dashed, dash pattern=on 1pt off 1pt, line width=0.3mm] (0, 0) -- (0, -0.25); 
        \draw[fill=black] (-0, -0.3) -- (0.1, -0.15) -- (-0.1, -0.15) -- cycle;
    \end{tikzpicture}

    \vspace{-2.5mm}
    
    \begin{tcolorbox}[boxrule=0pt, colback=blue!10]

    \footnotesize{
    \textbf{Q2:}
    Give me an answer of yes or no, can this technology be used in node.js, python, PHP, ruby, perl applications development, and building: "\{tag\_name\}, \{description\}"
    }

    \end{tcolorbox}
    
    \vspace{-2mm}
    
    \begin{tikzpicture}
        \draw[->, dashed, dash pattern=on 1pt off 1pt, line width=0.3mm] (0, 0) -- (0, -0.25); 
        \draw[fill=black] (-0, -0.3) -- (0.1, -0.15) -- (-0.1, -0.15) -- cycle;
    \end{tikzpicture}

    \vspace{-2.5mm}

    \begin{tcolorbox}[boxrule=0pt, colback=blue!10]

    \footnotesize{
    \textbf{Q3:}
    Give me an answer of yes or no, does this statement describe a technology that is related to the programming language itself rather than the application development and building: "\{tag\_name\}, \{description\}"
    }

    \end{tcolorbox}
    
    \caption{Desc-CoT Prompting.}
    \vspace{-2mm}
    \label{fig:desc_cot}
\end{figure}
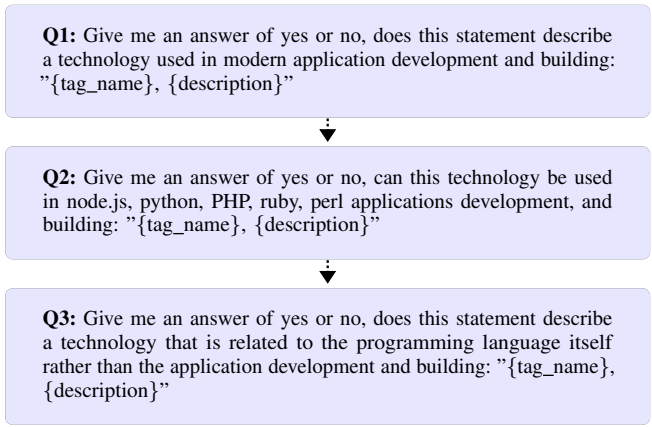

Figure~\ref{fig:desc_cot} illustrates the CoT approach. 
First, \textit{Q1} determines whether the tag and description describe a technology used in modern application development. 
Next, \textit{Q2} refines the selection to technologies supported by our target interpreted languages. 
Finally, \textit{Q3} filters out tags focusing on syntax issues. 
Note that the goal here is simply to reduce the number of tags of interest to us. 
SSCGuard utilizes ChatGPT-4o to process the CoT prompts. 
The filtered tags are then used to retrieve the corresponding questions, prioritizing those with the highest number of answers. 

\noindent{\textbf{Question Transformation.}}
Due to the overlap with other programming languages (e.g., Java) in the collected questions, SSCGuard then refines the question dataset by replacing non-interpreted languages. 
For example, the question "\textit{How to parse JSON in Java}" is transformed by replacing \textit{Java} with \textit{Nodejs}, \textit{Python}, \textit{PHP}, \textit{Ruby}, and \textit{Perl}, resulting in five new questions of the original question. 

Similarly, mainstream CI services (e.g., GitHub Actions and Jenkins) adopt different workflow syntaxes. 
To simplify the subsequent analysis, SSCGuards standardizes all CI service references to GitHub Actions by replacing mentions of other services. 
For example, the question "\textit{How to set environment variables in Jenkins?}" is transformed by replacing \textit{Jenkins} with \textit{GitHub Actions}.

\subsection{Response Generation}
SSCGuards first generates the collected questions into five sets of prompts, then feeds the prompts into the target LLMs to get responses. 

\subsubsection{Sets of Prompts}
Among the five prompt sets, four control sets focus on coding questions related to software \textit{development}, while one targets CI configuration for the \textit{build}. 
The four control sets are designed to investigate whether LLMs are more likely to suggest external components with potential security threats when the prompts clearly ask for them, such as requesting more external components.

\noindent{\textbf{Original Question (Q1).}}
This set uses plain questions as prompts and is labeled with the \textit{Q1}, serving as the baseline.

\noindent{\textbf{Code Examples (Q2).}}
This set refines the prompts by explicitly requesting the target LLMs to generate code examples to solve the original questions. 
These prompts, labeled \textit{Q2}, are designed to evaluate whether explicitly asking for code examples leads to an increased reference to external components compared to \textit{Q1}.

\noindent{\textbf{Single Package (Q3).}}
This set makes a more direct request based on the original questions, explicitly asking the target LLMs to provide a package and code example to solve the problems (labeled as \textit{Q3}).

\noindent{\textbf{Five Packages (Q4).}}
The last set refines the problems by asking the target LLMs to provide five packages that can be used to solve the original questions. 
This set is labeled as \textit{Q4}.

\noindent{\textbf{CI Prompt Set.}}
The prompt in this set is designed to be \textit{please give me a complete GitHub Actions configuration file to solve or realize the following problem or task, \{question\}}, where the \textit{\{question\}} is from the collected questions.

\subsubsection{Response Generation}
SSCGuard feeds the generated prompts to the four target LLMs. 
For ChatGPT-3.5-turbo and ChatGPT-4o-mini, SSCGuards employs the OpenAI batch API~\cite{open_batch_api}. 
For Llama-3.1-8b and Llama-3.1-sonar, SSCGuards utilizes the Perplexity API~\cite{perplexity_api}, which does not support batch processing. 
Therefore, it sends each prompt as an individual HTTPS request and records the responses for further analysis.

\subsection{Response Analysis}
In this part, SSCGuard first extracts the external component information from the responses generated by target LLMs, then feeds the extracted external components to perform threat detection to output results.

SSCGuards extracts the required external component information (e.g., recommended packages) from the responses generated by the target LLMs. 
For the \textit{original question} and \textit{code example} prompt sets, it extracts packages directly from the generated code (e.g., commands such as \texttt{npm install <package-name>}). 
In addition, it also extracts URLs hosting JavaScript libraries, CSS files, and third-party resources/services from the generated code. 
For the \textit{single package} and \textit{five packages} prompt sets, it extracts only the packages, as they are explicitly requested in the prompts provided to the LLMs. 

For the \textit{CI prompt set}, SSCGuards extracts the generated CI workflow configurations from the responses, and then stores them as independent configuration files (i.e., YAML files). 

\begin{figure*}[t]
    \centering
    \hspace{-10mm}
    \begin{minipage}{0.3\textwidth}
        \centering
        \includegraphics[scale=0.34]{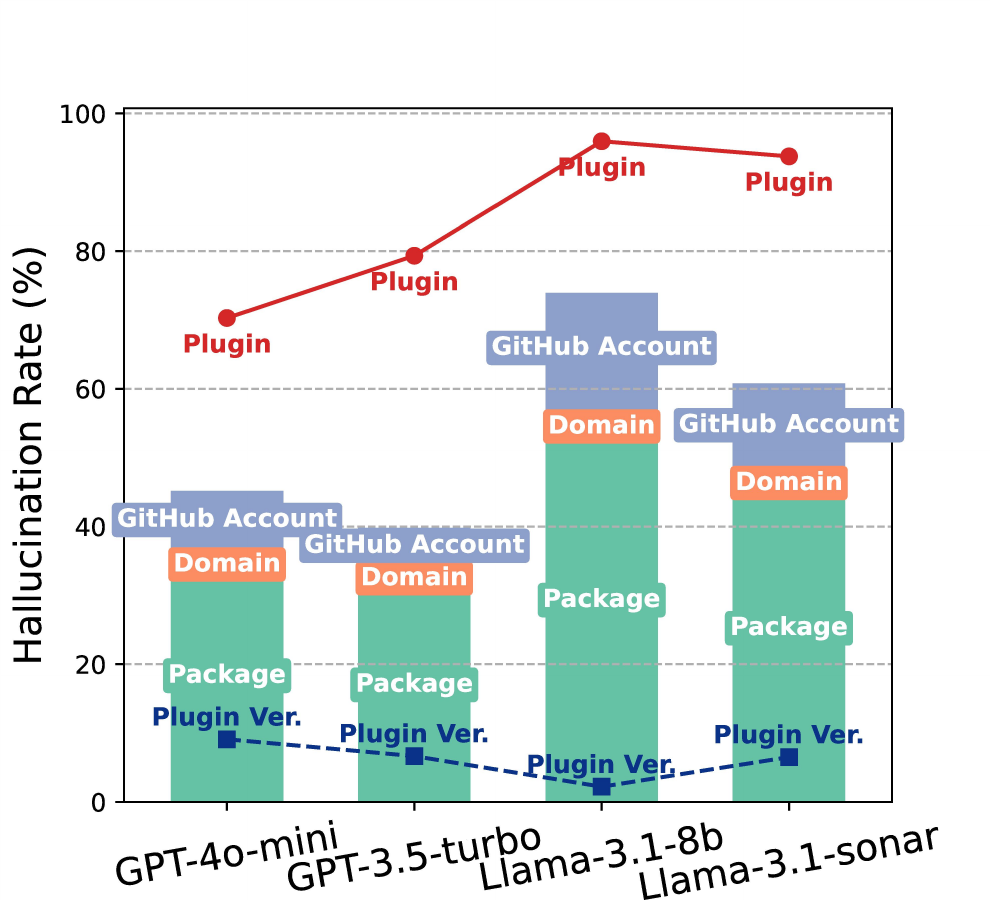}
        \captionsetup{font=small, labelfont={bf,small}, justification=centering, width=1\linewidth}
        \caption{Hallucination Rates across Four LLMs.}
        \vspace{-2mm}
        \label{fig:hallu_category}
    \end{minipage}
    \hspace{3mm}
    \begin{minipage}{0.3\textwidth}
        \centering
        \includegraphics[scale=0.34]{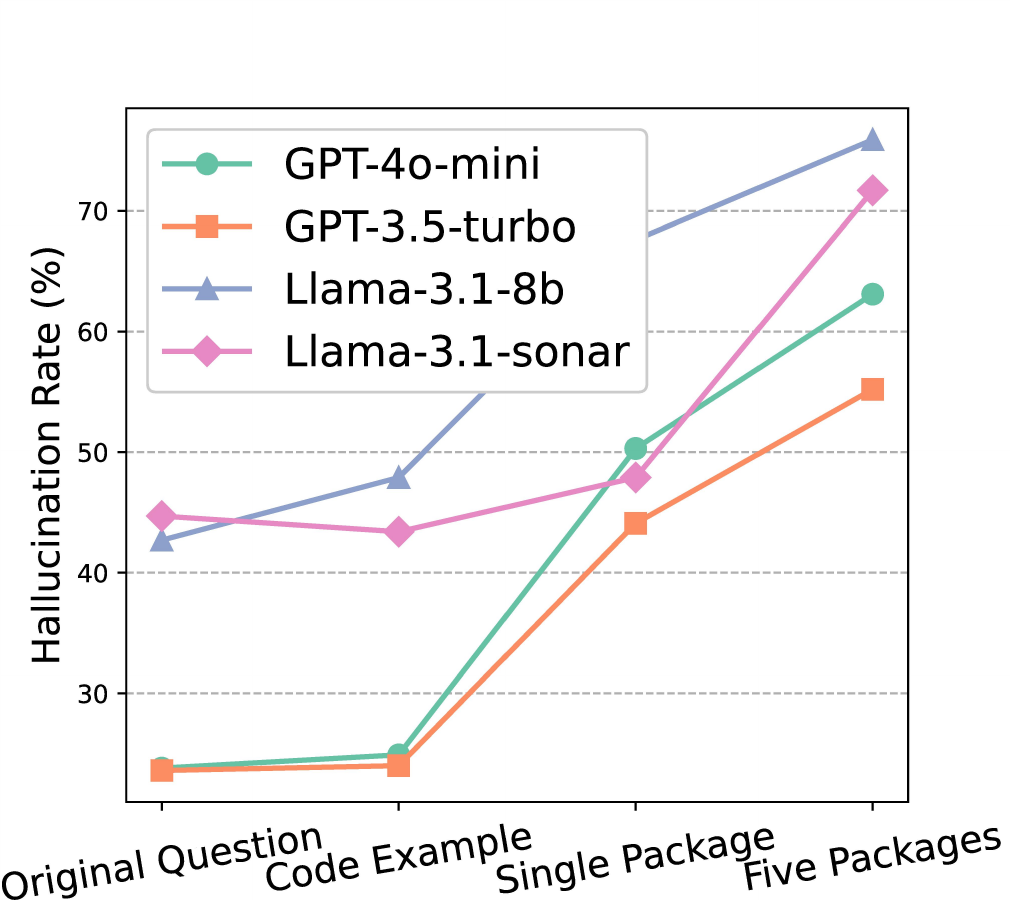}
        \captionsetup{font=small, labelfont={bf,small}, justification=centering, width=1\linewidth}
        \caption{Effect of Prompt Set on Hallucination Rates.}
        \vspace{-2mm}
        \label{fig:hallu_prompt_sets}
    \end{minipage}
    \hspace{5.7mm}
    \begin{minipage}{0.3\textwidth}
        \centering
        \includegraphics[scale=0.34]{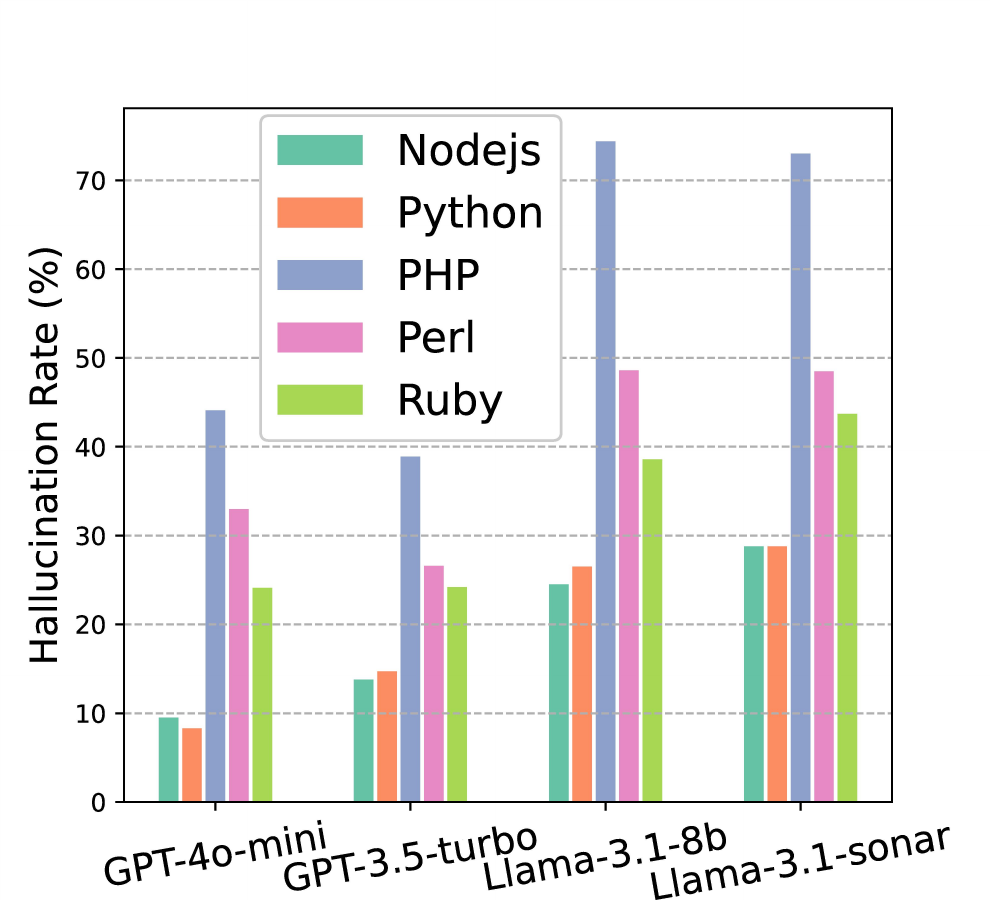}
        \captionsetup{font=small, labelfont={bf,small}, justification=centering, width=1\linewidth}
        \caption{Hallucination Rates Per Programming Language.}
        \vspace{-2mm}
        \label{fig:hallu_language}
    \end{minipage}
    
\end{figure*}

\subsection{Threat Detection}
\label{subsec:security_detector}

SSCGuards first verifies the status of the extracted external components (e.g., whether they exist or not) through HTTPS requests. 
For JavaScript libraries, CSS files, and third-party resources/services (e.g., GitHub repositories), it makes direct requests to their URLs, and records the returned status (e.g., \texttt{200} and \texttt{404}) for verification. 
For packages, it utilizes the corresponding registry APIs~\cite{npm_api, python_api, php_api, ruby_api, perl_api} to verify their status.

In the case of \textit{package hallucination}, it keeps the packages with the status of \texttt{404} (i.e., not found), suggesting these packages are available for publishing. 

For \textit{domain hallucination}, it keeps the URLs with \texttt{exception}, where an error occurs during the request process (e.g., exceeded max retries). 
\texttt{Exceptions} typically indicate that the server hosting the component is unreachable. 
After identifying these URLs, it queries GoDaddy to determine whether the domains are available for registration or purchase.  

For \textit{GitHub account hallucination}, it keeps the repositories with the status \texttt{404}.
Then it extracts their associated accounts and utilize the GitHub public API to check the accounts availability.

For \textit{vulnerable third-party resources/services versions}, it utilizes the \texttt{jsrepository-v4.json} file from \texttt{retire.js}~\cite{jsrepository_v4}, which contains a collection of the 64 most popular JavaScript libraries along with their corresponding versions and associated vulnerabilities. 

For \textit{third-party resources/services absence}, it keeps the requested URLs that return a status code of \texttt{404}, as this indicates the server itself is found, the requested object cannot be retrieved.

For \textit{package deprecation}, it first keeps the packages with the status code \textit{200}, indicating that they exist. 
Then, it utilizes official registry APIs~\cite{npm_api,php_api,perl_api} to determine whether the queried packages have been marked as deprecated.

SSCGuard analyzes the security issues in CI using the extracted configuration files. 
For code injection, it integrates ARGUS~\cite{muralee2023argus}, a static taint analysis tool, to detect code injection vulnerabilities in extracted configuration files.

Next, it extracts CI plugins and their specified versions from the extracted configuration files. 
These extracted plugins and versions are then utilized to analyze plugin-related issues. 

For \textit{CI plugin hallucination}, it utilizes the GitHub public API to check if the plugin exists. 
If the API returns a status code of \texttt{404}, it indicates that the plugin is hallucinated. 

For \textit{CI plugin redirection hijacking}, it first requests the plugins to check whether they are redirected to new locations. If redirection occurs, it then uses the GitHub public API to verify whether the old locations are available for registration.

It verifies the existence of the plugin's versions by checking both the tag list and branch name list of the plugin using the GitHub APIs. 
A version is considered a hallucination if it is not found in either the tag list or the branch list.
For \textit{CI plugin version reuse}, if the version exists in both the tag list and the branch list, it is identified as a tag-branch mix reuse. 

\subsection{Ethical Considerations}

We are committed to conducting our study with strict adherence to ethical considerations. 
To investigate security issues related to the SSC in LLM-generated code, we carry out a large-scale prompting of selected LLMs and analyze their responses. 
We also check the status of external components across relevant platforms (e.g., package registries). 
All actions are performed through official web interface, and no third-party accounts are compromised. 
All data used in our analysis is publicly available and obtained through legitimate methods. 
All experiments (e.g., CI configurations with code injection) are conducted using our own accounts.

\section{Evaluation}
\label{sec:evaluation_new}

\noindent{\textbf{Dataset.}}
SSCGuard collects 25,530 Stack Overflow tags with valid description and then obtains 3,791 required tags.
From these, we manually select the top 122 tags based on the question counts and further obtain their questions based on the number of answers.
Overall, SSCGuard obtains 104,782 coding questions, and 20,010 CI-related questions.
These questions are used to generate 419,128 prompts (for each question, four prompts are generated into four sets) and 20,010 prompts, respectively.

\noindent{\textbf{Large Language Models Settings.}} 
We use our framework and dataset to evaluate SSC hallucinations on four target LLMs.
GPT-4o-mini and GPT-3.5-turbo are proprietary models, thus, their internal settings are not publicly available. 
For the open-source models Llama-3.1-8b and Llama-3.1-sonar, we utilize the default settings of the Perplexity API, with a temperature of 0.2, a top\_p value of 0.9, a top\_k value of 0, and a customized max\_tokens limit of 10,000. 

\subsection{Quantitative Results}
Overall, we find that all identified SSC threats occur consistently in the code and CI configurations generated across all models. 
Below, we present some representative and notable results. 

\subsubsection{External Component Hallucination}

The bars in Figure~\ref{fig:hallu_category} show the average rate of hallucinated packages, domains, and GitHub accounts in four LLMs. 
Among them, package hallucination is the most prevalent, with an average rate of 33.01\% to 52.77\%, followed by GitHub account hallucination (5.83\% to 18.04\%). 
Domain hallucination is the least prevalent, ranging from 1.67\% to 3.35\%. 
The lines in Figure~\ref{fig:hallu_category} present the hallucination rates of the CI plugin and the version.

\noindent{\textbf{Package Hallucination.}}
The results in Figure~\ref{fig:hallu_category} reveal that hallucinated packages consistently occur for all programming languages on all target models. 

Figure~\ref{fig:hallu_prompt_sets} presents the average package hallucination rates across four sets on four LLMs.
The result shows a significant increase in hallucination rates for the \textit{single package} (Q3) and \textit{five packages} (Q4) sets, in which the prompts explicitly request the LLMs to provide package names.
For example, for GPT-4o-mini, the hallucination rate for Python is 8.36\% when responding to the \textit{original question}. 
However, this rate increases significantly to 37.09\% when responding to the \textit{single package} prompts. 
It seems that the LLMs often prioritize fulfilling the request by generating package names regardless of whether they actually exist. 

Figure~\ref{fig:hallu_language} displays the hallucination rates for five programming language types for the \textit{original question} prompt set. 
Nodejs and Python generally exhibit the lowest hallucination rates (9.5\% and 8.3\%). 
In contrast, PHP shows the highest hallucination rate, reaching 74.4\% for Llama-3.1-8b. 
Specifically, for PHP packages, the format follows the structure \textit{vendor/package-name}. 
If either \textit{vendor} or \textit{package-name} is hallucinated, the PHP package does not exist. 
LLMs frequently recommend PHP packages using only part of the required format, which is the main reason for PHP's high hallucination rate. 
For example, for GPT-4o-mini, the incomplete package rates are: 4.84\% for Q1, 3.58\% for Q2. 
The situation becomes much worse in Q3 and Q4, where the prompts explicitly ask LLMs to provide a package, with the rate of 61.09\% for Q3, 40.60\% for Q4. 

Additionally, we find hundreds of hallucinated Nodejs packages identified in CDN URLs vary across different models (24, 37, 184, 200 for GPT-4o-mini, GPT-3.5-turbo, Llama-3.1-8b, and Llama-3.1-sonar, respectively), which can also be hijacked by attackers. 

Across four LLMs, Llama-3.1-sonar and Llama-3.1-8b generally exhibit higher hallucination rates compared to GPT-4o-mini and GPT-3.5-turbo. 
For example, in the \textit{original question} prompt set, the package hallucination rates for Llama series are 42.7\% and 44.7\%, whereas for the GPT series, they are 23.8\% and 23.6\%. 

\noindent{\textbf{Domain Hallucination.}}
We find that hallucinated domains are a prevalent issue in the code generated by all four target LLMs. 
Most of these models have about 2-4\% domains that are hallucinated or dangling, which can be immediately taken over by attackers. 
For example, we find that the domain \texttt{camanjs.com}, referenced by GPT-4o-mini, used to be associated with the npm package \texttt{caman}, which still receives 433 weekly downloads at present. 
However, the domain \texttt{camanjs.com} has since expired and is available for purchase. 
Adversaries could take over the domain and hijack its residual traffic immediately.
Interestingly, Llama-3.1-8b shows great results with the \textit{original question} (0.32\%). 
However, when we explicitly ask it to recommend external components, performance drops significantly (4.58\% for the \textit{single package}).

\noindent{\textbf{GitHub Account Hallucination.}}
We consistently observe a lot of hallucinated and non-existent GitHub accounts.
Among these, Llama-3.1-8b demonstrates the highest average hallucination rate at 18.04\%, while GPT-3.5-turbo shows the lowest rate at 5.83\%. 
However, this could be attributed to the low number of accounts extracted from responses provided by GPT-3.5-turbo. 
Notably, Llama-3.1-8b recommends 6 repositories associated with the same account, \texttt{gradle-php}, which is a hallucinated account.

\noindent{\textbf{CI Plugin Hallucination.}}
The red line in Figure~\ref{fig:hallu_category} presents the CI plugin hallucination rates. 
Among extracted plugins, a significant proportion (from 70.29\% for GPT-4o-mini to 95.95\% for Llama-3.1-sonar) could not be found. 
These results are much worse than package hallucination. 
To understand the potential impact, we further investigate whether the associated GitHub accounts are vulnerable to hijacking using the GitHub API. 
We find that these non-existent plugins correspond to hundreds of GitHub accounts.
For all four models, about 15.44\% to 35.35\% are hijackable.  
Also, LLMs tend to recommend multiple plugins under the same account. 
For example, the GitHub account \texttt{k8s-github-actions} does not exist, while LLama-3.1-8b recommends 26 plugins under this account.

\begin{table}[t]
\small
\caption{Some Plugins with Hallucinated Versions.}
\label{ta:hallucination_versions}
    \centering
    \begin{tabular}{cccc}
        \toprule
        \textbf{LLM} & \textbf{Plugin} & \makecell{\textbf{Actual}\\\textbf{Version}} & \makecell{\textbf{Hallucinated}\\\textbf{Version}} \\
        \midrule
        \cellcolor[HTML]{EFEFEF}GPT-4o-mini & \cellcolor[HTML]{EFEFEF}\makecell{akhileshns/\\heroku-deploy} & \cellcolor[HTML]{EFEFEF}3 &  \cellcolor[HTML]{EFEFEF}23 \\
         \\
         GPT-3.5-turbo & \makecell{ad-m/github-\\push-action} & 4 & 7 \\ 
      \rowcolor[HTML]{EFEFEF} Llama-3.1-8b & \makecell{appleboy/\\ssh-action}  & 6 & 12  \\ 
      Llama-3.1-sonar & \makecell{appleboy/\\ssh-action} & 5 & 10 \\
        
        \bottomrule
    \end{tabular}
    \vspace{-3mm}
\end{table}

\noindent{\textbf{CI Plugin Version Hallucination.}}
Figure~\ref{fig:hallu_category} (blue line) also shows that hallucinated versions of plugins are quite common in LLM-generated configurations (from 2.22\% for Llama-3.1-8b to 9.12\% for GPT-4o-mini). 
Table~\ref{ta:hallucination_versions} shows some examples for each LLM with both the actual version and hallucinated versions. 
We can see that the hallucinated version is much larger than the actual one (e.g., 23 versus 3).

\noindent{\textbf{CI Plugin Version Reuse.}}
We find several cases (fewer than 20) in which the branch and tag of the referenced CI plugins share the same name.
For example, the plugin \texttt{microsoft/setup-msbuild} is recommended by both GPT-4o-mini and GPT-3.5-turbo, with versions \texttt{v1.0.1}, \texttt{v1.0.2}, and \texttt{v1.0.3} serving as both branch and tag names.
This indicates that LLMs might not be able to tell the differences between branches and tags when referencing CI plugins.

\subsubsection{Vulnerable External Components}
Table~\ref{table:overview_rest_risks} presents the average rates of identified vulnerable external components.

\noindent{\textbf{Vulnerable Third-party Resources/Services Versions.}} 
The results in Table~\ref{table:overview_rest_risks} suggest that it is common for LLMs to reference outdated versions. 
For example, GPT-4o-mini recommends a version \texttt{0.21.1} of the \texttt{axios} library, which is affected by a high-severity vulnerability~\cite{axios_cve}. 
Overall, the number of outdated library versions is similar for all models across different sets (from 13.33\% to 23.27\%).

\noindent{\textbf{Vulnerable CI Configuration. }}
The vulnerable configurations detected with code injection are very low (from 0.10\% to 2.18\%). 
The results show that, Llama models generate less vulnerable configuration files than GPTs. 
Particularly, one configuration generated by GPT-4o-mini is detected to contain 6 vulnerabilities, 4 of which are classified as high severity. 

\begin{table}[t]
\small
\caption{Results of Vulnerable Components.}
\label{table:overview_rest_risks}
    \centering
    \begin{tabular}{ccccc}
        \toprule
        \textbf{Threats} & \makecell{\textbf{GPT-4o}\\\textbf{-mini}} & \makecell{\textbf{GPT-3.5}\\\textbf{-turbo}} & \makecell{\textbf{Llama-}\\\textbf{3.1-8b}} & \makecell{\textbf{Llama-}\\\textbf{3.1-sonar}} \\
        \midrule
        \cellcolor[HTML]{EFEFEF}\makecell{Vuln 3rd-party\\Versions} & \cellcolor[HTML]{EFEFEF}5.52\% & \cellcolor[HTML]{EFEFEF}6.46\% & \cellcolor[HTML]{EFEFEF}6.97\% & \cellcolor[HTML]{EFEFEF}4.58\% \\
        \makecell{Vuln CI \\ Configurations} & 2.18\% & 1.29\% & 0.10\% & 0.12\% \\
        
        \bottomrule
    \end{tabular}
    \vspace{-3mm}
\end{table}

\subsubsection{Outdated External Components}
Table~\ref{table:overview_outdated} presents the average rates of outdated external components. 

\noindent{\textbf{Third-party Resources/Services Absence.}}
As shown in Table~\ref{table:overview_outdated}, the threat of absent third-party resources/services is quite common: all models show more than 10\% absent rate, and 23.27\% of components recommended by Llama-3.1-8b are missing.
By checking the details, we find that these external components mainly include JS libraries and CSS files.

\noindent{\textbf{Package Deprecation.}}
In Nodejs, PHP, and Perl, the registries provide mechanisms to mark packages as deprecated or abandoned, enabling the identification of deprecated packages by querying the registry APIs. 
We identify 2.59\%-4.69\% deprecated packages that are still valid but no longer maintained. 
It is worth mentioning that in the Nodejs packages, we also find a significant number of unpublished packages.
These unpublished packages are deleted from the registries by their owners.
There is an additional concern with unpublished packages referenced in the LLM-generated responses: their owners can re-publish these packages with totally different code inside. 

\noindent{\textbf{CI Plugin Redirection Hijacking.}}
We only find a few redirection plugins in target LLMs (from 0.37\% to 0.64\%). 
Our investigation reveals that all the accounts corresponding to the old locations still exist, suggesting that they are not vulnerable to hijacking at present.
Although they are not vulnerable, LLMs do recommend redirected plugins, which have potential risks.

\subsection{Qualitative Insights}

\subsubsection{Challenges.}
Our results demonstrate several SSC-related challenges for LLM coding.
First, LLMs might be \textit{trained on incomplete and often outdated data}, thus are prone to recommend outdated or unpublished packages, causing potentially severe SSC threats.
For example, GPT-4o-mini recommends a version \texttt{0.21.1} of the \texttt{axios} library, which is affected by a high-severity vulnerability~\cite{axios_cve}.

Another finding is that \textit{LLMs prioritize fluency over factual accuracy}, often “filling in the blanks” rather than verifying the existence of external components.
For example, our results show that, the package hallucination rates increase rapidly when LLMs are explicitly asked to recommend packages.

\subsubsection{Hallucinated Components Overlap Comparison.}
We calculate the overlap rate of the hallucinated components generated by four LLMs.
The data used for the calculation are obtained from the prompt sets \textit{original question} (Q1), \textit{code example} (Q2), and \textit{single package} (Q3).
Interestingly, the average overlap rates for hallucinated packages, GitHub accounts, and domains are 0.62\%, 0.98\%, and 0.70\%. 
Although hallucinated components are a common phenomenon in all four LLMs, the overlap rates are quite low. 
This might suggest that \textit{hallucinated components are model-specific}. 

However, the overlap rate between models from the same family (i.e., GPT series and Llama series) is much higher: the hallucinated package overlap rates for the GPT series and Llama series are 6.02\% and 11.13\%, respectively.
The reason could be that models within the same family may share similar (and outdated) training data.

\begin{table}[t]
\small
\caption{Results of Outdated External Components Across Four LLMs.}
\label{table:overview_outdated}
    \centering
    \begin{tabular}{ccccc}
        \toprule
        \textbf{Threats} & \makecell{\textbf{GPT-4o}\\\textbf{-mini}} & \makecell{\textbf{GPT-3.5}\\\textbf{-turbo}} & \makecell{\textbf{Llama-}\\\textbf{3.1-8b}} & \makecell{\textbf{Llama-}\\\textbf{3.1-sonar}} \\
        \midrule
        \cellcolor[HTML]{EFEFEF}\makecell{3rd-party R/S \\ Absence} & \cellcolor[HTML]{EFEFEF}16.91\% & \cellcolor[HTML]{EFEFEF}13.33\% & \cellcolor[HTML]{EFEFEF}23.27\% & \cellcolor[HTML]{EFEFEF}22.92\% \\
        \makecell{Package \\ Deprecation} & 3.71\% & 4.69\% & 2.59\% & 3.05\% \\
        \bottomrule
    \end{tabular}
    \vspace{-3mm}
\end{table}

\subsubsection{Semantic Similarity for Hallucinated Packages.} 
Using the set \textit{original question}, we analyze the semantic relationship between the prompt and the recommended packages. 
We aim to investigate whether LLMs primarily focus on a specific word in the prompt and fabricate package names accordingly. 
We use BERT to extract the semantics of both hallucinated package names and individual words within the prompt. 
Then we use cosine similarity to quantify the semantic similarity between each word in the prompt and the corresponding hallucinated package. 
If there exists a value of cosine similarity that exceeds 0.6, it indicates a meaningful semantic relationship.
The semantic similarity results range from 51.52\% to 56.79\% across the four LLMs, demonstrating that \textit{LLMs indeed tend to focus on a single word when fabricating package names}. 

Notably, we identify instances where the semantic similarity value reached 1 (5.13\% to 11.33\%), indicating that LLMs directly use a word from the prompts as package names. 
The result suggests that adversaries could systematically collect a large number of prompts and utilize LLMs to generate package names based on those prompts, thereby hijacking non-existent packages.

\subsubsection{Package Overgeneralization}
Overgeneralization in LLMs refers to the tendency of the model to produce content that is overly generic and lacks nuance. 
Using the set \textit{original question}, we analyze whether this phenomenon occurs in code generation. 
In particular, we investigate how LLM overgeneralization is reflected in package recommendations.
We define LLMs recommending overly broad packages as package overgeneralization, for example, suggesting the npm package \texttt{package-name} (i.e., \texttt{npm install package-name}). 
While \texttt{package-name} is a valid and installable package, its description reveals that it is actually intended for a brunch application~\cite{package-name}. 
Therefore, we manually examined all overly broad packages recommended by LLMs for the set \textit{original question}. 
The overgeneralization numbers range from 5 to 22 across the four target LLMs, indicating that \textit{package overgeneralization does occur during code generation.}

\label{sec:defences}

\begin{figure}
    \centering

    \begin{tikzpicture}[baseline]

        \node[anchor=east, align=right, fill=gray!20, inner sep=6pt] at (0, 0) {\footnotesize \textbf{1. Extract Packages}};
        
        \node[anchor=west] at (0, 0) {
            \begin{tcolorbox}[width=0.32\textwidth, boxrule=0pt, colback=blue!10, left=2mm, right=2mm, top=0.5mm, bottom=0.5mm]
            \footnotesize{"\{baseline\_code\} Are there any packages installed using commands in the content above? If yes, please extract all the packages and their registries from the content above. Do not alter package names, and output packages and registries."}
            \end{tcolorbox}
        };
    \end{tikzpicture}

    \vspace{-1mm}
    \hspace{2.5cm}
    \begin{tikzpicture}
        \draw[->, dashed, dash pattern=on 1pt off 1pt, line width=0.3mm] (0, 0) -- (0, -0.25); 
        \draw[fill=black] (0, -0.3) -- (0.1, -0.15) -- (-0.1, -0.15) -- cycle;
    \end{tikzpicture}

    \vspace{-1mm}

    \begin{tikzpicture}[baseline]

        \node[anchor=east, align=right, fill=gray!20, inner sep=6pt] at (0, 0) {\footnotesize \textbf{2. Confirm Package} \\ \footnotesize \textbf{Existence}};
        
        \node[anchor=west] at (0, 0) {
            \begin{tcolorbox}[width=0.32\textwidth, boxrule=0pt, colback=blue!10, left=2mm, right=2mm, top=0.5mm, bottom=0.5mm]
            \footnotesize{"Please only provide a yes or no answer to whether the package '\{package\_name\}' exists in the '\{registry\}' registry. Ensure that you check for the existence of the exacted package name. Do not alter the package name, and do not mark it as existing based on similar package names."}
            \end{tcolorbox}
        };
    \end{tikzpicture}

    \vspace{-1mm}
    \hspace{2.5cm}
    \begin{tikzpicture}
        \draw[->, dashed, dash pattern=on 1pt off 1pt, line width=0.3mm] (0, 0) -- (0, -0.25); 
        \draw[fill=black] (0, -0.3) -- (0.1, -0.15) -- (-0.1, -0.15) -- cycle;
    \end{tikzpicture}

    \vspace{-1.5mm}

    \begin{tikzpicture}[baseline]
        \node[anchor=east, align=right, fill=gray!20, inner sep=6pt] at (0, 0) {\footnotesize \textbf{3. Re-generate Code}};
        
        \node[anchor=west] at (0, 0) {
            \begin{tcolorbox}[width=0.32\textwidth, boxrule=0pt, colback=blue!10, left=2mm, right=2mm, top=0.5mm, bottom=0.5mm]
            \footnotesize{"Please use the \{packages\} to solve or realize the following problem or task: \{original\_question\}. Only use the provided packages. Do not alter package names, and do not add additional packages."}
            \end{tcolorbox}
        };
    \end{tikzpicture}
    \caption{The Design of Chain-of-Confirmation.}
    \vspace{-2mm}
    \label{fig:coc}
\end{figure}
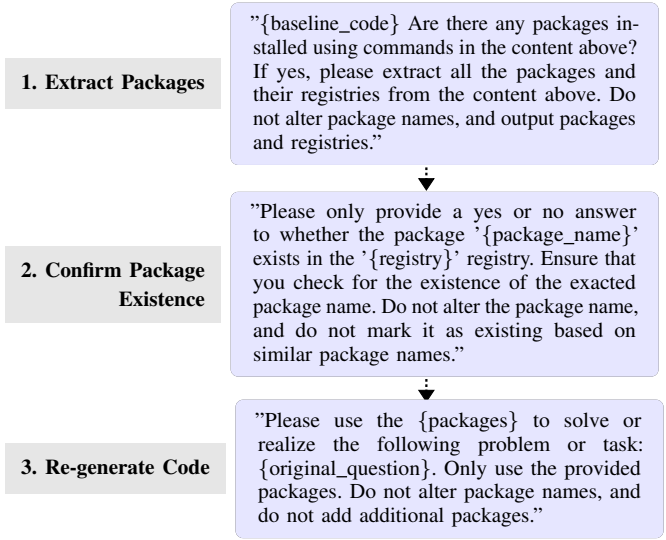

\section{Mitigation Discussions}

Extensive research efforts~\cite{zhang2024llm,dhuliawala2023chain,ji2023towards} have been devoted to understanding and mitigating hallucination in LLMs, while fully addressing it still remains challenging. 
To specifically mitigate the identified SSC-related risks in LLM-generated code, we propose two defensive practices: prompt-based defense and middleware-based defense. 
The prompt-based defense enables developers to craft prompts meticulously, hence mitigating the identified issues in LLM-generated code. 
However, the effectiveness of this approach is inherently limited by the root causes of LLM hallucination in code generation, such as reliance on outdated library documentation and incorrect usage examples~\cite{zhong2024can, zhang2024llm}. 
To address this gap, we further propose the middleware-based defense as a more robust strategy. 
This defense involves deploying our tool, SSCGuard as an intermediary layer between the LLM and developers. 
By analyzing the generated code, SSCGuard can identify problematic external components (e.g., failed URL requests) and inform developers.

\subsection{Prompt-based Defense}

Using package hallucination as a representative example, we explore strategies to reduce hallucinations in LLM-generated code.

\noindent\textbf{Naive Approaches.} We first demonstrate that naive approaches may not work well. 
We design two \textit{\textbf{single}}-prompt mechanisms aimed at reducing package hallucination. 
For the first approach, the prompt simply asks the LLM to \textit{\textbf{not}} recommend URLs and packages that can not be successfully requested.
For the second approach, the prompt asks the LLM to \textit{\textbf{avoid}} recommending URLs and packages.

We randomly select a subset of questions from the set \textit{code example} (\textit{Q2}), containing 20,721 coding questions.  
We evaluate these questions and mitigation using GPT-4o-mini. 

Table~\ref{ta:prompt_based_defense} shows the results. For the original questions (i.e., baseline), we extract 1,925 packages from LLM-generated code. Among them, 146 packages are hallucinated, with a rate of 7.48\%. 
For the first single-prompt strategy, we can only extract 1,398 packages, more than 500 packages fewer than the original one. 
Among them, 90 packages are hallucinated, with a rate of 6.43\%. 
The second single-prompt strategy can reduce the hallucination rates to 5.68\%, but can only extract 809 packages, less than half of the original ones. 
The results show that, naive approaches (e.g., a single prompt) can only slightly reduce the occurrence of hallucination, while also largely reducing the number of recommended packages.

\noindent{\textbf{Our Approach.}}
We propose a novel method, Chain-of-Confirmation (CoC), to reduce hallucination for the referenced packages, while maintaining a similar number of recommended packages (as the original question).
The idea is to first let LLMs confirm the existence of referenced packages, and then re-generate the code.
The rationale is that, LLMs tend to provide more accurate facts with an independent verification question, compared with the original question~\cite{dhuliawala2023chain}.
Thus, an extra confirmation step asking the LLMs to confirm the existence of recommended packages should reduce package hallucination.
Figure~\ref{fig:coc} illustrates the details of CoC. 
We first instruct the LLM to extract the packages and their corresponding registries from the LLM-generated code. 
Next, we instruct the LLM to confirm the existence of the extracted packages within their respective registries, and return a binary result: "Yes" or "No". 
Finally, we ask the LLM to re-generate the code for the given problem using only the packages whose existence has been confirmed.

CoC can extract 1,569 packages, which is slightly less than the original question, but more than both single-prompt approaches. 
Among them, only 57 packages are hallucinated, with a rate of 3.63\% of the total packages. This hallucination rate is less than half of the original question. 
The results show that CoC can effectively reduce the package hallucination rate, while ensuring that the number of recommended packages is not significantly reduced.

\begin{table}[t]
\small
  \caption{Performance of Prompt-Based Defense Practices.}
  \label{ta:prompt_based_defense}
  \centering
  \begin{threeparttable}
    \begin{tabular}{l@{}lll@{}} 
 \toprule
      \multirow{2}{*}{\textbf{}} & \textbf{Extracted~}  & \textbf{Hallucinated~} & \textbf{Hallucination~} \\ 
      & \textbf{Packages~} & \textbf{~Packages~} & \textbf{~Rates} \\
     \midrule[0.5pt]
     \rowcolor[HTML]{EFEFEF} Baseline & 1,925 & 146 & 7.48\% \\ 
     Single-prompt I & 1,398 & 90 & 6.43\% \\ 
     \rowcolor[HTML]{EFEFEF} Single-prompt II $~$ & 809 & 46 & 5.68\% \\ 
     \makecell{Chain-of-\\Confirmation} $~$ &  1,569 & 57 &  \textbf{3.63\%} \\ 
     \bottomrule
    \end{tabular}
  \end{threeparttable}
  \vspace{-3mm}
  \end{table}

\subsection{Middleware-based Defense}

For the second defense mechanism, we propose to modify and deploy \textit{Response Analysis} and \textit{Threat Detection} as an intermediary layer between the LLMs and developers. 
Such middleware can be implemented as a browser plugin. 
Basically, SSCGuards takes LLM-generated code as input, and extracts external components and their information referenced within the code.
Then, the middleware collects various information from official resources (e.g., the existence of packages, actual license) as detailed in Section~\ref{subsec:security_detector}, and detects corresponding threats. 
For example, for CI risk analysis, the middleware can integrate ARGUS~\cite{muralee2023argus} to detect the potential code injection vulnerabilities in the workflow configurations generated by LLMs. 
Then, it can extract all plugins and their specified versions referenced within the LLM-generated workflows. 
The proposed middleware further queries the GitHub public APIs to verify the status of these plugins and their specified versions. 
If risks are identified, the middleware presents the analysis results, along with the LLM-generated code, to the developers as security alerts.

\label{sec:related_work}

\section{Related Work}

\noindent\textbf{SSC Package Security.}
Extensive research efforts have been focused on understanding the vulnerabilities within the SSC and developing strategies to enhance its security~\cite{ladisa2023sok, lin2024untrustide, zahan2022weak, zahan2023software}. 
During the \textit{development} phase, Duan et al.~\cite{duan2020towards} introduced a comparative framework capable of identifying hundreds of malicious packages within popular interpreted programming language package registries, including PyPI, npm, and RubyGems. 
Gu et al.~\cite{gu2023investigating} investigated twelve package-related security threats across six software registries, including \textit{package hijacking attacks} and \textit{package redirection hijacking attacks}, a vulnerability caused by deleted third-party website accounts (e.g., GitHub).
Additionally, several studies showed that vulnerable packages can be exploited to compromise their dependent packages, creating cascading security risks~\cite{hejderup2015dependencies, decan2019empirical}. 
While these previous works motivate our research, this paper explores a different dimension: investigating how LLM hallucinations can cause exploitable packages and third-party accounts.

\noindent\textbf{Dangling References.} 
Security risks caused by various dangling resources have also been studied recently~\cite{liu2016all,borgolte2018cloud,squarcina2021can,lauinger2017game,gruss2018use,mariconti2017s}. 
Hao et al.~\cite{hao2013understanding} observed that spammers frequently re-register expired domains. 
Lever et al.~\cite{lever2016domain}  identified that the malicious re-registration of expired domains is the root cause of many security issues.  
In contrast, our study focuses on domains hallucinated by LLMs, which present a new and unique threat as they can be opportunistically registered by adversaries.
In addition, Lauinger et al.~\cite{lauinger2018thou} conducted a comprehensive study of client-side JavaScript library usage, revealing that 37\% of websites include at least one outdated library with a known vulnerability. 
Our study further sheds light on a distinct problem: LLMs might recommend outdated and vulnerable JavaScript libraries, even when the latest versions are available.

\noindent\textbf{CI Security.} Additionally, many studies have been devoted to the security of CI services during the \textit{build} phase~\cite{li2022robbery, koishybayev2022characterizing,gu2023continuous,gu2024more}. 
There are two closely related works. Li et al.~\cite{li2024toward} explored seven attack vectors targeting several mainstream CI platforms. Particularly, they disclosed the plugin redirection hijacking threat and CI plugin version reuse threat, which are also discussed in this paper.
Muralee et al.~\cite{muralee2023argus} developed ARGUS, a static taint analysis tool designed to detect code injection vulnerabilities in GitHub Actions workflows.
Our research also relies on ARGUS to examine potential code injection vulnerabilities in LLM-generated CI workflows.

\noindent\textbf{Security of LLM Generated Code.}
Many recent research works have studied the quality of the LLM-generated code~\cite{liu2024exploring,liu2024no,zhang2024effectiveness}, including security~\cite{jesse2023large, li2024enhancing, tambon2024bugs}, usability~\cite{liu2024refining}~\cite{nguyen2022empirical}~\cite{vaithilingam2022expectation}, and correctness~\cite{liu2024no}~\cite{yeticstiren2023evaluating}. 
For example, Pearce et al.~\cite{pearce2022asleep} conducted experiments on GitHub Copilot to generate code in high-risk scenarios, showing that many programs that Copilot completes are vulnerable. 
Additionally, Sandoval et al.~\cite{sandoval2023lost} demonstrated that AI-assisted coding with LLMs tends to introduce more critical bugs when programming low-level languages such as C.
Zhang et al.~\cite{zhang2024effectiveness} evaluated the effectiveness of state-of-the-art LLMs, revealing that LLMs tend to to produce invalid workflows and workflows prone to code injection vulnerabilities. 
However, none of these works examine SSC-related security risks, and our work is the first to largely investigate SSC-related risks in LLM-generated code.
A parallel work~\cite{spracklen2024we} investigates package hallucinations in LLM-generated code. 
However, within the broader context of software supply chain security, packages represent only one class of potentially hallucinated components. 
In contrast, our work conducts a systematic and comprehensive study of hallucinations across a wider range of external components referenced by LLMs.

\noindent\textbf{General LLM Security.}
Finally, many research works have focused on the security of LLMs, such as data poisoning attacks~\cite{aghakhani2024trojanpuzzle, schuster2021you, yan2024llm}, jailbreak attacks~\cite{wei2024jailbroken, shen2024anything, deng2023jailbreaker}, and prompt injection attacks~\cite{liu2023prompt, liu2024formalizing, greshake2023not}. 
Unlike these types of works, our work does not attempt to break LLMs, but focuses on how LLMs can be abused for SSC-related security risks.

\section{Conclusion}
This paper investigates SSC-related security threats caused by LLM inherent issues. 
We have examined three categories of threats, with eleven potential threats, on source code and CI configuration files. 
We have designed a tool called SSCGuards, and conducted a large scale analysis with 439,138 prompts on four target LLMs.
Results show that the identified SSC-related threats consistently exist in all target LLMs, suggesting that the developers should not blindly trust LLMs for software development. 
We have also discussed mitigation mechanisms.






%

\bibliographystyle{unsrt}
\bibliography{ref}

\end{document}